# An organometallic chimie douce approach to new Re$_x$W$_{1-x}$O$_3$ phases

Christian Helbig, Rudolf Herrmann, Franz Mayr, Ernst-Wilhelm Scheidt, Klaus Tröster, Jan Hanss, Hans-Albrecht Krug von Nidda and Wolfgang Scherer*

**Re$_x$W$_{1-x}$O$_3$·H$_2$O and Re$_x$W$_{1-x}$O$_3$ phases are prepared by a new organometallic *chimie douce* concept employing the organometallic precursor methyltrioxorhenium.**

Crystalline WO$_3$ has found widespread interest as electrochromic material, applied *e.g.* as thin film for "smart windows",[1] and as gas sensor.[2] At room temperature monoclinic γ-WO$_3$ (P2$_1$/n) represents the thermodynamically most stable phase and consists of tilted WO$_6$ octahedra. Its structure is therefore related via group - subgroup relationships with cubic ReO$_3$ (*Pm-3m*), which represents the aristotype of the BO$_3$ perovskite family. Accordingly, it should be possible to form solid solutions of mixed Re$_x$W$_{1-x}$O$_3$ phases due to the structural resemblance of both parent oxides and the similar ionic radii of Re$^{VI}$ and W$^{VI}$.[3] While WO$_3$ represents an insulator, ReO$_3$ displays metallic behaviour with a specific conductivity in the same range as crystalline copper.[4] Hence, solid solutions of both oxides might lead to new phases with interesting electronic properties like the related Na$_x$WO$_3$ phases, which are benchmark systems to study chemically-induced metal-to-insulator transitions.[5] However, up to now mixed Re$_x$W$_{1-x}$O$_3$ phases were only accessible in small quantities under extreme conditions at high pressure and high temperatures (65 kbar, 1200°C).[6] Here, we propose a new organometallic *chimie douce* method which allows to synthesise Re$_x$W$_{1-x}$O$_3$ phases in large quantities by a low temperature process at ambient pressure.

A convenient aqueous synthesis of WO$_3$·yH$_2$O (y = 1, 2, 1/3) from tungstates *via* tungstic acid has been reported and discussed earlier.[7-9] We found that mixed hydrated WO$_3$/ReO$_3$ phases are formed from sodium metatungstate containing the Keggin ion [H$_2$W$_{12}$O$_{40}$]$^{6-}$ and the organometallic precursor, CH$_3$ReO$_3$, (<u>me</u>thyl<u>tri</u>ox<u>o</u>rhenium; MTO) in diluted HCl at 100°C.[10] The new Re$_x$W$_{1-x}$O$_3$·H$_2$O phases are obtained as dark green to black microcrystalline platelets, in contrast to WO$_3$·H$_2$O which shows a bright yellow colour when prepared under the same conditions (Fig. 1). The choice of the precursors is crucial; no mixed WO$_3$/ReO$_3$ phases are obtained from sodium tungstate Na$_2$WO$_4$·2H$_2$O which consists of isolated WO$_6$ octahedra. We interpret these observations as a kinetic effect: acidification of sodium tungstate initiates a fast condensation process[8] leading to WO$_3$·H$_2$O precipitates. This process, however, is apparently delayed when α-metatungstate is used instead as precursor. In this case, hydrolysis of the [H$_2$W$_{12}$O$_{40}$]$^{6-}$ ion to release the neutral monooxo precursor [WO(OH)$_4$(OH$_2$)] provides the sufficient time span for the preconditioning of our organometallic precursor (*e. g.* CH$_3$ReO$_2$(OH)$_2$ formation)[11] to take part in the following oxolation process. Once incorporated, MTO looses the methyl group, accompanied by reduction of Re$^{VII}$ to Re$^{VI}$.[11] We note that only pure WO$_3$·H$_2$O is precipitated when other inorganic Re$^{VII}$ species like perrhenate salts are added to the reaction mixture instead of MTO. Hence, the salient capability of our organometallic precursor to form Re$^{VI}$ oxide species by methane elimination in water appears to be crucial for the successful tungsten/rhenium substitution.[12] The composition and homogeneity of the samples was ascertained by ICP, SEM-EDX, X-ray diffraction and microanalysis up to a Re content of 12%. For *x* larger than 0.12 phase separation via ReO$_3$ formation is observed.

\* wolfgang.scherer@physik.uni-augsburg.de

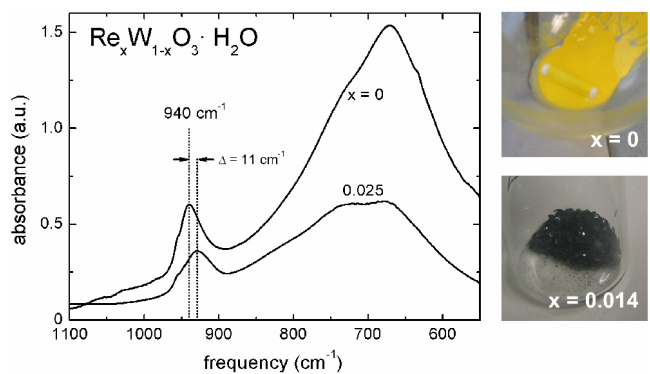

**Fig. 1** IR spectra (KBr) of WO$_3$·H$_2$O and Re$_x$W$_{1-x}$O$_3$·H$_2$O with *x* = 0.025. The right hand side shows photographs of tungstite and a mixed phase (*x* = 0.014).

IR spectroscopic studies (KBr pellets; Fig. 1) show a characteristic band at 940 cm$^{-1}$ in the parent tungstite, WO$_3$·H$_2$O, which shifts to lower frequency in our new Re$_x$W$_{1-x}$O$_3$·H$_2$O samples with increasing rhenium content. This mode can be assigned by the stretching mode of the W=O bond (see structural motif in Fig. 5) which appears to become softened by stepwise substitution of W(d$^0$) by the less Lewis-acidic Re(d$^1$) centres.[9,13] Hence, substitution of only 2.5% of tungsten atoms by rhenium lowers the corresponding W/Re=O streching frequency by 11 cm$^{-1}$. Stretching vibrations of the O-W-O or mixed O-W/Re-O units give rise to a broad signal around 700 cm$^{-1}$ (680 and 730 cm$^{-1}$ have been reported for undoped tungstite[13]) which appears to be less effected by the substitution of tungsten *vs.* Re.

A further indication of electronic changes induced by the rhenium content of the mixed phases can be found in the dehydration behaviour as studied by TGA using a temperature ramp from 25 to 230°C during 45 minutes. As shown in Fig. 2, all phases loose a total of one molecule of water per formula unit (7%), but the ease of dehydration depends on the rhenium content. While significant water loss (5% of the total loss) of WO$_3$·H$_2$O starts only above 180°C,[9] the mixed phases with *x* = 0.025 (0.053) of rhenium content loose water already at 140°C (90°C). Hence, replacement of the W(d$^0$) atoms by the more electron-rich and less Lewis-acidic Re(d$^1$) centres facilitates the dehydration of water molecules which are interconnecting corner-sharing WO$_6$ layers via hydrogen bonding. The samples with rhenium contents *x* > 0.12, which show phase separation, loose water in two clearly distinguishable steps at different temperatures. The inflection points

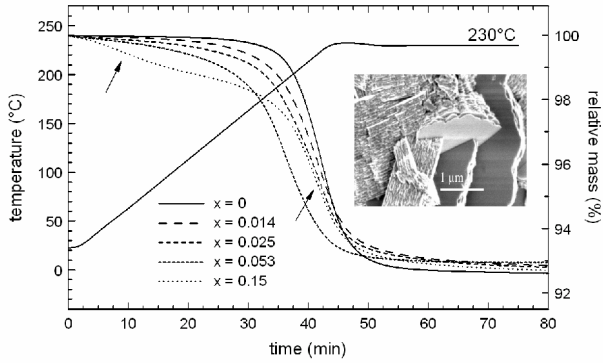

**Fig. 2** TGA dehydration curves of selected $Re_xW_{1-x}O_3 \cdot H_2O$ phases (right axis) and the respective temperature profile (left axis). The inset displays a typical SEM image for the sample $x = 0.014$ revealing a morphology typical for layered compounds.

of the dehydration curve of the sample with 15% rhenium content are marked by arrows in Fig. 2.

Magnetisation measurements were performed for all $Re_xW_{1-x}O_3 \cdot H_2O$ samples presented in this paper. In Fig. 3 the inverse susceptibility $\chi^{-1}(T) = B/M$ is pictured, representatively for $x = 0.025$. Below 100 K $\chi(T)$ is well accounted for by a Curie-Weiss law $\chi(T) = C/(T-\theta) + \chi_0$ with a marginal itinerant contribution $\chi_0 \approx 0.03$ memu/mol and $\theta \approx 0$ K. The vanishing Curie-Weiss temperature $\theta$ indicates no correlations between the residual localised $d^1$ electrons at the Re atoms. The effective paramagnetic moment $\mu_{eff} = 0.06\ \mu_B$ obtained from the Curie constant $C$ indicates that 0.1% of the metal sites carry a $d^1$ moment, which correspond to an amount of 4% of the Re atoms. In the inset of Fig. 3 the magnetisation $M(B)$ at 2 K is displayed. The fitted Brillouin function (solid line) takes into account a $d^1$ state with a quenched orbital moment ($L = 0$). The saturation magnetisation $M_{sat}$ is proportional to the quantity of the aligned local spins. The observed small saturation value of $M_{sat} \approx 10^{-3}\ \mu_B$ / f.u. corresponding to an amount of 0.1% $d^1$ moments at the metal sites corroborates our susceptibility results (for a single $d^1$ moment $M_{sat} = 1\ \mu_B$ / f.u.).

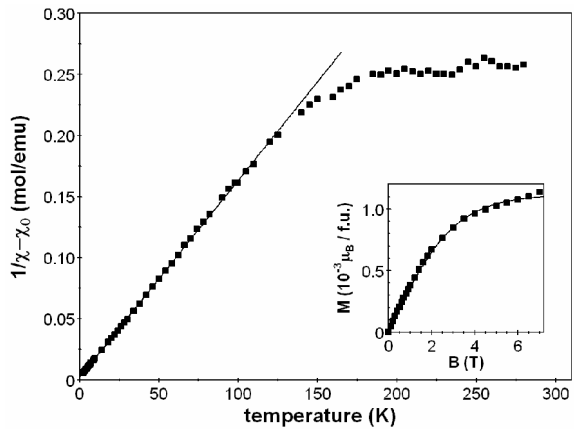

**Fig. 3** The inverse magnetic susceptibility $\chi^{-1}$ of $Re_xW_{1-x}O_3 \cdot H_2O$ ($x = 0.025$) measured in an external magnetic field $B = 1$ T. A diamagnetic core electron contribution $\chi_{Dia} = -68\ \mu$emu/mol was subtracted from the data. The solid line is a linear fit with a Curie-Weiss law. The inset displays the field dependence of the magnetic moment $M$ per formula unit at 2 K together with a fitted Brillouin function (solid line).

As one would expect from the marginal $\chi_0$ value, the resistivity $\rho$ of pressed powder samples of $Re_xW_{1-x}O_3 \cdot H_2O$ is generally high (about 10 $\Omega$cm) and reveals an insulating temperature dependence ($d\rho/dT < 0$).

ESR spectra (9.35 GHz, 4 K) confirm the presence of localised $d^1$ moments at the Re atoms: a signal at $g = 2$ is observed, showing a hyperfine coupling (six lines) to rhenium ($I = 5/2$ for $^{185}$Re and $^{187}$Re), with an average coupling constant $A \sim 620$ Oe (0.058 cm$^{-1}$; see Fig. 4).[14]

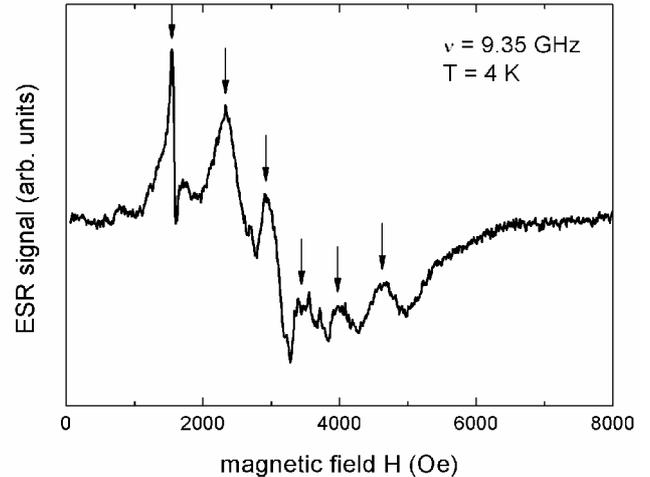

**Fig. 4** An ESR spectrum of $Re_xW_{1-x}O_3$ ($x = 0.025$) at 4 K. The six lines indicate the hyperfine coupling of electron spins localised at $Re(d^1)$ centres.

Finally, a strong indication for the change in the $WO_3 \cdot H_2O$ lattice due to rhenium doping is found by X-ray powder diffractometry. The mixed hydrated phases are isotypic to the orthorhombic $WO_3 \cdot H_2O$ tungstite structure (*Pmnb*). Fig. 5 (b) shows an overlay of the diffraction pattern of $WO_3 \cdot H_2O$ and the phase with 2.5% rhenium content. Both patterns confirm the high crystallinity of our samples and do not hint for any parasitic phase. Since tungsten and rhenium atoms have about the same scattering cross-section for X-rays and due to the isotypic relationship between the pure $WO_3 \cdot H_2O$ and mixed $Re_xW_{1-x}O_3 \cdot H_2O$ the position and therefore possible ordering of the Re and W atoms could not be disclosed yet.

However, closer inspection of the diffraction pattern (Fig. 5 (a)) reveals that some Bragg peaks of the rhenium substituted samples are slightly but significantly shifted to smaller angles (Bragg peaks (002) and (022) in Fig. 5 (a)) in comparison with the pattern of the parent compound $WO_3 \cdot H_2O$. Accordingly, Rietveld refinements reveal a widening of the $WO_3 \cdot H_2O$ lattice by 1.5 pm in $c$ direction (from 5.1264(4) Å to 5.1416(1) Å) after Re doping ($x = 0.025$).

Analysis of the diffraction pattern of dehydrated $Re_xW_{1-x}O_3 \cdot H_2O$ samples, however, was complicated due to the partial amorphous character of the products (Fig. 6 (a)). The resulting pattern closely resembles the ones found earlier for dehydrated $WO_3 \cdot H_2O$ phases which were reported to represent the hitherto unknown metastable form of cubic $WO_3$.[15] However, a direct comparison with the cubic diffraction pattern of highly crystalline $Re_{0.25}W_{0.75}O_3$ powder obtained by bulk synthesis at 7.5 GPa and 1100°C ($a = 3.7516(2)$ Å; Fig. 6 (c))[16] clearly reveals that the new $Re_xW_{1-x}O_3$ phases obtained by dehydration should be indexed rather by a monoclinic than by a cubic unit cell. Precise lattice parameters for the new

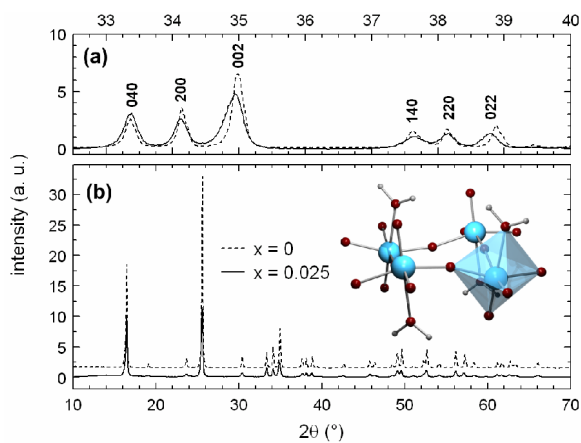

**Fig. 5** (b) X-ray powder diffraction pattern of WO$_3$·H$_2$O (tungstite) and Re$_x$W$_{1-x}$O$_3$·H$_2$O ($x = 0.025$). (a) The upper part is an enlargement of the diffraction range between 32 and 40°. The inset shows the structural motif of tungstite representing layers of corner-sharing polyhedra which are interconnected in the third dimension along the crystallographic $a$ axis via W=O⋯H-O-H⋯O=W hydrogen bonding.

monoclinic Re$_x$W$_{1-x}$O$_3$ phase ($x = 0.025$, $a = 7.3165(4)$ Å, $b = 7.5131(4)$ Å, $c = 7.6911(5)$ Å, $\beta = 90.521(6)°$, $P2_1/n$) could be obtained by Rietveld analysis of the annealed samples (720°C, 1 day) (Fig. 6 (b)). Furthermore, a single crystal diffraction study on Re$_{0.02}$W$_{0.98}$O$_3$ ($a = 7.305(3)$ Å, $b = 7.534(2)$ Å, $c = 7.691(2)$ Å, $\beta = 90.88(3)°$) which was alternatively synthesised by chemical transport methods at 1000°C clearly reveals the isotypic relationship between monoclinic Re$_x$W$_{1-x}$O$_3$ phases obtained by *chimie douce* and classical ceramic methods.

Hence, at this stage we can demonstrate that we succeeded to obtain pure Re$_x$W$_{1-x}$O$_3$ phases by an organometallic *chimie douce* approach. An organometallic compound, methyltrioxorhenium, was found to be the only successful precursor suitable for the rhenium doping of WO$_3$ at low temperatures and ambient pressure. Diffraction studies reveal that the new Re$_x$W$_{1-x}$O$_3$ phases obtained by our *chimie douce* approach are not cubic but isotypic to monoclinic $\gamma$-WO$_3$. We could hence demonstrate that cubic phases of mixed Re/W trioxides still appear to remain a domain of high pressure high temperature methods. As a result of our studies a fast and simple *chimie douce* pathway to mixed Re/W trioxides has been opened as an alternative to ceramic routes which warrants further exploitation with respect to electronic design (*e.g.* electrochromic and gas sensing properties) and chemical behaviour (*e.g.* intercalation chemistry, oxidation catalysis).

Christian Helbig, Rudolf Herrmann, Franz Mayr, Ernst-Wilhelm Scheidt, Klaus Tröster, Jan Hanss, Hans-Albrecht Krug von Nidda and Wolfgang Scherer
*Institut für Physik, Universität Augsburg, Universitätsstr. 1, D-86135 Augsburg, Germany. Fax:+49-(0)821-598-3227; Tel:+49(0)821-598-3350; E-mail: wolfgang.scherer@physik.uni-augsburg.de*

**Acknowledgement:**
This work was supported by the SFB 484 of the Deutsche Forschungsgemeinschaft (DFG).

# Notes and references

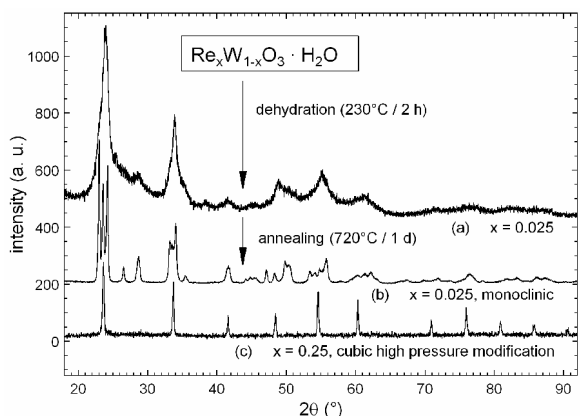

**Fig. 6** (a) X-ray powder diffraction pattern of dehydrated Re$_x$W$_{1-x}$O$_3$·H$_2$O ($x = 0.025$) and (b) the annealed sample; (c) diffraction pattern of the cubic high pressure modification of Re$_{0.25}$W$_{0.75}$O$_3$.